\documentclass[aps,showpacs,preprint,groupedaddress,floatfix]{revtex4}
\usepackage{graphicx}
\usepackage{dcolumn}
\usepackage{bm}

\begin{document}

\title{Ground state properties of rare-earth nuclei in the relativistic
      Hartree-Bogoliubov model with density-dependent meson-nucleon couplings}
\author{T. Nik\v si\' c}
\author{D. Vretenar}
\affiliation{Physics Department, Faculty of Science, University of Zagreb, 
Croatia, and \\
Physik-Department der Technischen Universit\"at M\"unchen, D-85748 Garching,
Germany}
\author{G.A. Lalazissis}
\affiliation{Department of Theoretical Physics, Aristotle University of
Thessaloniki, GR-54124, Greece}
\author{P. Ring}
\affiliation{Physik-Department der Technischen Universit\"at M\"unchen, 
D-85748 Garching,
Germany}
\date{\today}

\begin{abstract}
The relativistic mean-field effective interaction with 
density-dependent meson-nucleon couplings DD-ME1 is tested 
in the calculation of deformed nuclei.
Ground-state properties of six isotopic chains ($60\le Z \le 70$) in the
region of rare-earth nuclei are calculated by using the
relativistic Hartree-Bogoliubov (RHB) model with the DD-ME1 
mean-field interaction, and with the Gogny D1S force
for the pairing interaction.
Results of fully self-consistent RHB calculations 
for the total binding energies,
charge isotope shifts and quadrupole deformation parameters are compared
with the available empirical data.
\end{abstract}

\pacs{21.10.Dr, 21.10.Ft, 21.30.Fe, 21.60.Jz}
\maketitle

\bigskip \bigskip

Relativistic mean-field (RMF) models have been very successfully employed in 
analyses of a variety of nuclear structure phenomena, not only in nuclei along 
the valley of $\beta$-stability, but also in exotic nuclei with extreme isospin 
values and close to the particle drip lines. The relativistic Hartree-Bogoliubov
(RHB) model~\cite{Gon.96}, based on the relativistic mean-field theory and on the
Hartree-Fock-Bogoliubov framework, provides a unified description of mean-field
and pairing correlations, which is particularly important for the structure of
very weakly bound nuclei close to the particle drip lines.
An interesting alternative to the highly successful RMF models
with nonlinear meson self-interaction terms, is an effective 
hadron field theory with medium dependent
meson-nucleon vertices. Such an approach retains the basic
structure of the relativistic mean-field framework, 
but can be more directly related to the
underlying microscopic description of nuclear interactions.
In particular, the density-dependent relativistic hadron field 
(DDRH) model~\cite{FLW.95} has been successfully applied
in the calculation of nuclear matter and ground-state properties of 
spherical nuclei~\cite{TW.99}, and extended to
hypernuclei~\cite{KHL.00}, neutron star matter~\cite{HKL.01a}, and
asymmetric nuclear matter and exotic nuclei~\cite{HKL.01}. 
In Ref.~\cite{NVFR.02} 
we have extended the relativistic Hartree-Bogoliubov model to include
density-dependent meson-nucleon couplings. The effective Lagrangian is
characterized by a phenomenological density dependence of the $\sigma$,
$\omega$ and $\rho$ meson-nucleon vertex functions, adjusted to properties
of nuclear matter and finite nuclei. The DD-ME1 effective interaction has been
introduced and tested in the analysis of the equations of state for symmetric
and asymmetric nuclear matter, and of ground-state properties of the Sn and Pb
isotopic chains. It has been shown that, when compared to 
results obtained with standard nonlinear
relativistic mean-field effective forces, the DD-ME1 interaction  
provides an improved description of
asymmetric nuclear matter and of ground-state properties of $N \neq Z$ nuclei.
In Ref.~\cite{NVR.02} we have also shown that 
the relativistic random phase approximation (RRPA) with the 
DD-ME1 effective interaction, reproduces the experimental excitation
energies of multipole giant resonances in spherical nuclei.

Relativistic density-dependent effective mean-field interactions have 
never before been used in the calculation of deformed nuclei.
Of course, the structure of deformed nuclei presents
an important test for every effective interaction. Ground-state properties, 
in particular, are sensitive to the isovector channel of effective interaction, 
to the spin-orbit term of the effective 
single-nucleon potentials, and to the effective mass. For example,
in Ref.~\cite{LRR.99} the standard NL3 
nonlinear meson-exchange interaction \cite{LKR.97} has been employed in
a detailed RMF analysis of ground state properties of 1315 even-even nuclei
($10\le Z\le 98$), and it has been shown that  
this interaction produces very good results for deformed nuclei.

In this short note we test the 
DD-ME1 effective interaction in the region of rare-earth nuclei. We compare 
predictions of the RHB model for the total binding energies, charge isotope 
shifts, and ground-states quadrupole deformations of six 
even-Z isotopic chains 
($60\le Z \le70$), with available empirical data. 
The DD-ME1 effective interaction is used in the particle-hole ($ph$)
channel, and pairing correlations are described by the pairing part of the
finite range Gogny D1S interaction~\cite{Gog.84}. The RHB equations are solved
self-consistently, with potentials determined in the mean-field approximation
from solutions of Klein-Gordon equations for the meson fields. The
Dirac-Hartree-Bogoliubov equations and the Klein-Gordon equations are solved by
expanding the nucleon spinors and the meson fields in terms of eigenfunctions of
a deformed axially symmetric oscillator potential~\cite{TRG.90}. The number of
oscillator shells in the expansion is 12 for nucleon fields, 
and 20 for meson fields.

The predictions of the RHB model for the total binding energies of the Nd, Sm,
Gd, Dy, Er and Yb isotopes are shown in Fig.~\ref{figA}, in 
comparison with the
experimental data~\cite{AW.95}. We notice a very good agreement over
the entire region of rare-earth nuclei. The maximum 
deviation of the calculated binding
energies is below $0.1$\% for all isotopes, except $^{142}$Nd, $^{144}$Sm, 
$^{146}$Gd, $^{148}$Dy and $^{150}$Er. For these nuclei 
the deviation from experimental binding energies is $0.2$\% .  

In Fig.~\ref{figB} we compare the theoretical values for the charge isotope
shifts with the data from Ref.~\cite{NMG.94}. The charge density is
obtained by folding the calculated point-proton density distribution with the
Gaussian proton-charge distribution. For the latter an rms radius of
0.8 fm is used, and the resulting ground-state charge radius reads
\begin{equation}
r_c = \sqrt{r_p^2+0.64}\quad {\rm fm} \;,
\end{equation}
where $r_p$ is the radius of the point-proton density distribution. 
The isotope shifts are
calculated with respect to a reference nucleus in each isotopic chain
\begin{equation}
\delta r_{ch}^2 = r_{ch}^2 - r_{ch}^2({\rm ref.})\;.
\label{shift}
\end{equation}
The reference nucleus is the $N=82$ isotope, except for the chains 
Dy and Yb, for which the reference nuclei are $^{156}$Dy and 
$^{168}$Yb, respectively.
The calculated charge radii reproduce in detail 
the empirical isotope shifts. A slight deviation from the empirical 
trend is observed only for heavier Dy nuclei.
However, even for $^{164}$Dy the deviation of the theoretical 
charge radius from the empirical value is only $0.3$\%. 

The ground-state quadrupole and hexadecupole 
deformation parameters $\beta_2$ and $\beta_4$
are calculated according to the prescription of Ref.~\cite{LQ.82}.
The theoretical values of the quadrupole deformation 
parameters are displayed in
Fig.~\ref{figC}, in comparison with the empirical data from Ref.~\cite{RNT.01}.
We notice that the RHB results reproduce not only the 
global trend of the data, but also the saturation of quadrupole deformation
for heavier isotopes.

In conclusion, we have applied the RHB model with the density-dependent
meson-nucleon couplings to the analysis of ground-state properties of six
isotopic chains ($60\le Z \le 70$) in the region of rare-earth nuclei. The 
DD-ME1 effective
interaction has been used in the $ph$-channel, and pairing correlations have
been described by the pairing part of the finite range Gogny D1S interaction.
An excellent agreement has been obtained in comparison with empirical
data on total binding energies, charge isotope shifts and
quadrupole deformation parameters. These results show that relativistic 
mean-field interactions with explicit density dependence of the 
meson-nucleon couplings, and in particular the   
DD-ME1 effective interaction, provide an accurate description of 
the structure of deformed nuclei.

\bigskip \bigskip
\leftline{\bf ACKNOWLEDGMENTS}

This work has been supported in part by the Bundesministerium
f\"ur Bildung und Forschung under project 06 TM 193, and by the
Gesellschaft f\" ur Schwerionenforschung (GSI) Darmstadt.

\newpage

\begin{figure}
\caption{The binding energies of Nd, Sm, Gd, Dy, Er and Yb isotopes,
       calculated in the RHB model with the DD-ME1 interaction, are compared
       with experimental data \protect\cite{AW.95}.}
\label{figA}
\end{figure}

\begin{figure}
\caption{Charge isotope shifts of Nd, Sm, Gd, Dy, Er and Yb isotopes.
       The results of the RHB calculation with the DD-ME1 effective interaction,
       and with the Gogny D1s interaction in the pairing channel,
       are compared with empirical data \protect\cite{NMG.94}.}
\label{figB}
\end{figure}

\begin{figure}
\caption{Comparison between the DD-ME1 predictions for the ground-state
	quadrupole deformation parameters of the Nd, Sm, 
       Gd, Dy, Er and Yb isotopes, and experimental
       data \protect\cite{RNT.01}}
\label{figC}
\end{figure}

\end{document}